\documentclass[aps,pra, twocolumn,superscriptaddress,longbibliography]{revtex4-2}
\usepackage{dsfont}
\usepackage{amsfonts}
\usepackage{array}
\usepackage{braket}
\usepackage{upgreek}
\usepackage{graphicx}
\usepackage{float}
\usepackage[margin=0.68in]{geometry}
\usepackage[dvipsnames]{xcolor}
\usepackage{derivative}
\usepackage{xcolor}
\usepackage{siunitx}
\usepackage{soul}
\usepackage{minitoc}
\usepackage{color}
\usepackage{amssymb}
\usepackage{amsmath, scalerel, breqn}
\usepackage{colortbl}
\usepackage{bbm}
\usepackage{physics}
\usepackage{bbold}
\usepackage[normalem]{ulem}
\usepackage{comment}
\usepackage{enumitem}
\usepackage[T1]{fontenc}
\usepackage{booktabs}

\renewcommand{\arraystretch}{1.2}

\definecolor{dark-red}{rgb}{0.84,0.15,0.15}
\definecolor{dark-blue}{rgb}{0.15,0.15,0.4}
\definecolor{medium-blue}{rgb}{0,0,0.5}
\definecolor{copper}{rgb}{0.72, 0.45, 0.2}
\definecolor{lightgreen}{RGB}{204, 255, 204}

\usepackage{hyperref}
\hypersetup{
    colorlinks, linkcolor={dark-red},
    citecolor={dark-blue}, urlcolor={medium-blue}
}

\usepackage[most]{tcolorbox}
\usetikzlibrary{patterns}
\pgfdeclarepatternformonly{editstrikeoutpattern}{\pgfqpoint{-1pt}{-1pt}}{\pgfqpoint{11pt}{11pt}}{\pgfqpoint{10pt}{10pt}}%
{
  \pgfsetlinewidth{0.4pt}
  \pgfpathmoveto{\pgfqpoint{0pt}{0pt}}
  \pgfpathlineto{\pgfqpoint{10.1pt}{10.1pt}}
  \pgfusepath{stroke}
}
\newtcolorbox{tcbstrikeout}{breakable,
 enhanced jigsaw,
 opacityback=0,
 parbox=false,
 boxrule=0mm,
 top=0mm,bottom=0pt,left=0pt,right=0pt,
 boxsep=0pt,
 frame hidden,
 finish={\fill[pattern=editstrikeoutpattern] (frame.north west) rectangle (frame.south east);}
}

\usepackage[capitalize]{cleveref}

\definecolor{green}{rgb}{0.08,0.7,0.05}

\usepackage{quantikz}
\soulregister\cite7
\soulregister\ref7

\newcommand{\subhead}[1]{\vspace{2mm}\par\noindent\textbf{#1}\par\vspace{1mm}\noindent}

\makeatletter
\def\@fnsymbol#1{\ensuremath{\ifcase#1\or *\or \dagger\or \dagger\or
   \mathsection\or \mathparagraph\or \|\or **\or \dagger\dagger
   \or \ddagger\ddagger \else\@ctrerr\fi}}
\makeatother

\renewcommand{\thetable}{\arabic{table}}

\begin{document}
\doparttoc 
\faketableofcontents 
\part{}

\setcitestyle{super}

\title{Quantum–Classical Separation in Bounded-Resource Tasks \\ Arising from Measurement Contextuality}
\author{Google Quantum AI and Collaborators\,$^\dagger$}

\begin{abstract} 
Quantum phenomena can be exploited to tackle certain problems beyond the reach of classical approaches. Here, we focus on coordinated decision-making among many players without communication, the study of which on current quantum processors has remained elusive. Using a superconducting qubit processor, we show that quantum contextuality enables certain tasks to be performed with success probabilities beyond classical limits. With a few qubits, we illustrate quantum contextuality with the magic square game, and quantify it through a Bell--Kochen--Specker inequality violation. %
To examine many-body contextuality, we implement the N-player GHZ game and separately solve a 2D hidden linear function problem, {advancing the state of the art on fidelity and system size in both tasks}. Our work introduces benchmarking of quantum processors using contextuality-based algorithms.
\par\medskip\noindent\textit{Published in Science} \textbf{394}, 870--874 (2026), DOI: \href{https://doi.org/10.1126/science.aec3042}{10.1126/science.aec3042}.
\end{abstract} 

\maketitle 

It is widely believed that distinct non-classical features allow certain quantum algorithms to outperform their classical counterparts, an expectation that has motivated extensive research on quantum processors. Yet proving separations between computational algorithms is extremely challenging; to date, no unconditional proof of quantum advantage exists for a general computational model. Nevertheless, a few rare algorithms demonstrate a provable separation in computational scaling—and therefore a clear advantage—under resource constraints beyond standard memory and time. On the hardware side, however, there is still no direct experimental measurement of the quantum–classical separation for these algorithms, nor of the degree of ‘quantumness’ in the corresponding many-body states, as sufficiently large quantum processors have only recently become available.

Beyond quantum–classical separations, a central question is which distinct aspect of quantum phenomena constitutes a computational resource. Although entanglement is well understood as a resource for quantum communication, some studies suggest that contextuality could be the key resource enabling the superiority of quantum computation.\cite{Raussendorf_PRA_2013,howard2014contextuality,bravyi2018quantum,Bravyi2020NatPhys,coudron2021trading,watts2019exponential,grier2020interactive,kretschmer2025} The role of contextuality in many-body computation, unlike in few-body tasks, remains poorly understood. Most studies on the information-theoretic aspects of many-body quantum states have focused on entanglement, with only a few exceptions examining contextuality.\cite{Deng2012,Tura2014, Schmied2016, Pelisson2016, Tura2017, bravyi2018quantum, Daniel2021} Quantum contextuality has been experimentally tested in  systems with a few quantum degrees of freedom\,\cite{kirchmair2009state, Bartosik:2009PRL,Amselem:2009PRL, moussa2010testing,Zhang_PRL2013,Hu_PRL_2016,Jerger2016contextuality,Wei2010contextuality,Leupold_PRL2018,van2019multipartite,Holweck2021,wang2022significant,CinelliPRL2005,YangPRL2005,BarbieriPRA2005,Aolita:2012PRA,Xu_PRL_2022,Laghaout_2022,Xue_PRL_2023,fabbrichesi2025tests,kelleher2025empirical}, and has seen limited exploration in many-body systems.

Quantum measurement contextuality arises from the observation that in certain quantum systems, the outcomes of measurements cannot be consistently pre-assigned in a way that is independent of the measurement context\,\cite{Spekkens2008,Abramsky2011, Budroni:2022RMP}. The discussion of this phenomenon traces back to the famous EPR argument and Einstein’s question, “\textit{Is the moon there when nobody looks?}”,\cite{Pais_RevModPhys1979,Mermin_Moon_1985} reflecting the realist belief that physical properties exist objectively and independently of observation. This notion of pre-existing properties became central to the EPR (Einstein–Podolsky–Rosen) argument for the incompleteness of quantum theory\,\cite{EPR_1935}, a view that was ultimately challenged by experimental tests.\cite{Freedman1972,Aspect1982}

 A set of measurements on a tripartite quantum state $\Psi = (|000\rangle - |111\rangle)/\sqrt{2}$ can illustrate the impossibility of assigning pre-defined values to physical observables\,\cite{Mermin1990,GHZ_1990}. When measuring the 3-qubit Pauli operators \(A=X_1Y_2Y_3\), \(B=Y_1X_2Y_3\), and \(C=Y_1Y_2X_3\), where $X_i$ and $Y_i$ denote the Pauli operators acting on qubit $i$, the outcomes are always +1. One can measure $A$, for example, by simultaneously measuring $X_1$, $Y_2$, and $Y_3$, obtaining outcomes $x_1$, $y_2$, and $y_3$, respectively, which are each $\pm 1$, such that product $x_1 y_2 y_3$  is always +1. However, a measurement of $D = X_1 X_2 X_3$, yields $-1$ even though multiplying $x_1 y_2 y_3 = +1$, $y_1 x_2 y_3 = +1$, and $y_1 y_2 x_3 = +1$ yields $x_1 x_2 x_3 = +1$, which one would expect to agree with the measurement of $D$ if one assumed that all of the $x_i$ and $y_i$ had pre-determined values. 
 This counterintuitive finding highlights the subtle role of measurement context in describing what is observed in quantum systems. 

\subhead{Magic square game}
Quantum contextuality-based tasks provide a way to exploit classically-impossible correlations to design co-operative games where players can win despite being unable to communicate during game play\,\cite{Cabello_PRL_2001_1,Cabello_PRL_2001,aravind2002,brassard2003multi,Aravind:2004AJP,brassard2005quantum}. In these so-called quantum pseudo-telepathy games, players achieve perfect winning by sharing an entangled quantum state and conditioning their strategies on measurement outcomes, instead of using a classical set of pre-agreed instructions. To experimentally illustrate contextuality, we implement  the magic square game\,\cite{Mermin_PRL_1990,PERES_1990,Cabello_PRL_2001_1,Cabello_PRL_2001,aravind2002} on our superconducting qubit processor. This is a cooperative game featuring two players, Alice and Bob, and a referee. The rules of the game are as follows.

\begin{figure*}[th!!]
\centering
\includegraphics[width=0.99\textwidth]{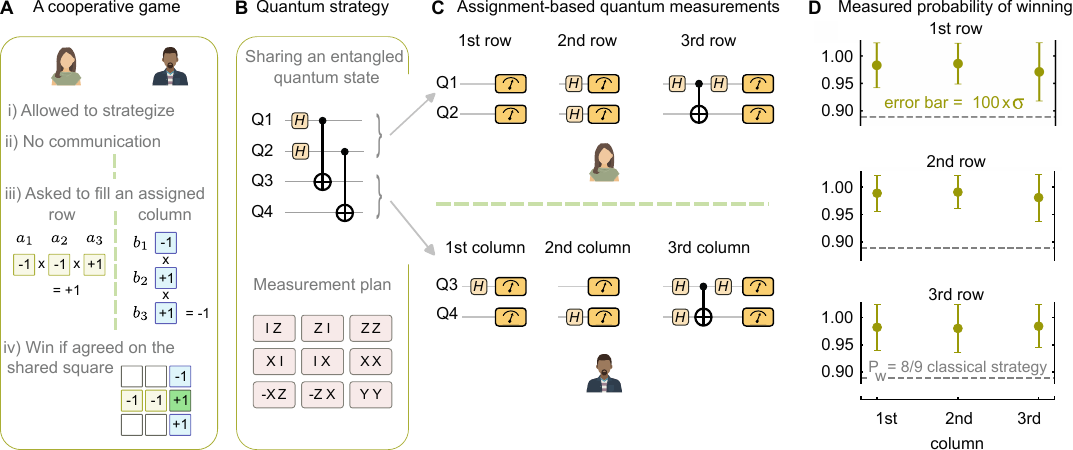}
\caption{\small{\textbf{The magic square game.} \textbf{(A)}, The rules of this game. The green dashed line indicates the incommunicado constraint. Alice gets a row and Bob a column of a $3 \times 3$ table; entries must satisfy row products $= +1$, column products $= -1$, and they win if their shared cell matches.\,\textbf{(B)} Leveraging a quantum strategy, the players share two Bell pairs and agree on what measurements to perform for each assigned row (Alice) or column (Bob), using the measurement plan table, where first Pauli in each cell refers to their first qubit.\,\textbf{(C)} Each player performs two simultaneous measurements and, given the product constraint rule (iii), they infer the third value needed to fill out their assigned row or column.\,\textbf{(D)} Measured probability of winning the game when played on our processor, shown for 9 distinct games (row or column assignment) sorted by where the assigned rows and columns intersect. The average probability of winning is $P_w=0.9830 \pm 0.0001$. Each game is played $N_s=100{,}000$ times and the error bar plotted for each game is 100 times the statistical uncertainty derived from the binomial distribution.}} 
\end{figure*}

\renewcommand{\theenumi}{\roman{enumi}}
\begin{enumerate}
    \item \textit{Resource sharing}: Initially, players are allowed to communicate and share a strategy. 
    \vspace{-2mm}
    \item \textit{Imposing non-communication}: {Alice and Bob are separated, and the referee randomly assigns to Alice the $j$-th row and to Bob the $k$-th column. They reply with three numbers each, $a_1$, $a_2$, $a_3$ and $b_1$, $b_2$, $b_3$ $\in \{\pm1\}$, respectively.}    
    \vspace{-2mm}
    \item \textit{Constraint}: {The product of the three values in Alice’s row assignments must be $+1$ whereas the product of the three values in Bob’s column assignments must be $-1$. That is,}
    \begin{equation}
       R_j=a_1a_2a_3=+1, \quad C_k=b_1b_2b_3=-1.
    \end{equation} 
    \vspace{-3mm}
    \item \textit{Winning condition}: They win if they have the same number at the intersecting cell. {That is, $a_k=b_j$}.
\end{enumerate}

Note that the row and column assignments are generated in each run, and not pre-determined values. Any classical strategy that they could share before they know their row and column assignment can be represented as a $3 \times 3$ table of the values $\pm 1$. In this table, the number $-1$ would need to appear an even number of times for each row and an odd number of times for each column. Therefore, the total number of $-1$ needs to be even and odd at the same time, which is impossible.  Thus there is no classical strategy that wins every time.  It can be seen that the best classical strategy results in a winning probability of 8/9, under the assumption that the referee assigns the players rows and columns uniformly at random. Here, we implement a quantum pseudo-telepathy strategy that enables players to win the game with probability $P_w=1$. Starting with a pair of Bell pairs,  

\begin{equation}
   |\Psi\rangle = \Big(\frac{|00\rangle +|11\rangle}{\sqrt2} \Big)_{Q1,Q3}\otimes \Big(\frac{|00\rangle +|11\rangle}{\sqrt2}\Big)_{Q2,Q4} ,
\end{equation}

\noindent Alice and Bob each take half of each Bell pair, i.e. Alice takes $Q1$ and $Q2$, and Bob takes $Q3$ and $Q4$. After they are separated and informed about their row and column assignment, they each use the pre-agreed Table 1 to select which measurements they should make. To play the game, we use a $2{\times}2$ square of qubits. We prepare Bell pairs between the vertical pairs and then identify the top two qubits as Alice and the bottom two as Bob. 

\begin{table}[htbp]
  \centering
  \setlength{\tabcolsep}{8pt}
  \renewcommand{\arraystretch}{1.2}

  \begin{tabular}{lccc}
    & \textbf{column 1} & \textbf{column 2} & \textbf{column 3} \\
    \textbf{row 1} & $I_1Z_2$      & $Z_1I_2$      & $Z_1Z_2$   \\
    \textbf{row 2} & $X_1I_2$      & $I_1X_2$      & $X_1X_2$   \\
    \textbf{row 3} & $-X_1Z_2$  & $-Z_1X_2$  & $Y_1Y_2$   \\
  \end{tabular}

  \caption{Measurement plan for the magic square game. The Mermin--Peres table is a set of measurement instructions that the two players can use to deterministically win the game, $P_w=1$, if given noiseless quantum resources. Here $I$, $X$, $Y$, and $Z$ denote the single-qubit identity and Pauli operators, with subscripts labeling the qubit on which each operator acts.}
  \label{tab:example_table}
\end{table}

\begin{figure*}[th!!]
\centering
\includegraphics[width=0.99\textwidth]{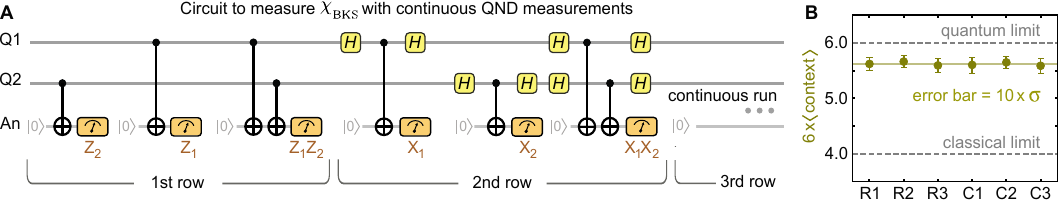}
\caption{\small{\textbf{Quantifying contextuality by measuring $\chi_{\text{BKS}}$.} \textbf{(A)}, Measuring $\chi_{\text{BKS}}$ requires successive quantum non-demolition\,(QND) measurements of two qubits\,(Q1, Q2) which we achieve by using an ancilla qubit (An). We construct a long circuit of 180 randomly selected rows or columns (contexts, 540 QND measurements) and run the sequence for $1{,}000$ times. Using single qubit and two qubit entangling gates, a random sequence of rows and columns of Table \ref{tab:example_table} is measured and after each measurement the ancilla is reset to the $|0\rangle$ state. \textbf{(B)}, Measuring each row or column (a ``context") entails three successive measurements, each $\pm 1$, which in the absence of noise, would multiply to +1 for the rows or $-1$ for the columns. Plotted are $6\langle R_i\rangle$ and $-6\langle C_i\rangle$, for all rows and columns; error bars indicate 10 $\times$ the statistical uncertainty, $\sigma$. The horizontal dashed line corresponding to $\chi_{\rm BKS}=4$ is the classical limit assuming pre-existing classical values of $\pm 1$ for each of the 9 elements of the table and the olive, horizontal solid line corresponds to the sums of the averages $\chi_{\text{BKS}} = 5.590 \pm 0.008$, as given in Eq.~\ref{eq:chi}. Additional data are provided in Figs.~S4 and S5\,\cite{SM}.}}   
\end{figure*}

In Fig.\,1D we show the result of playing the 9 games, sorted based on the intersection of the assigned row and column. Each of these distinct games is played $N_s=100{,}000$ times, yielding an average winning probability of $P_w=0.9830\pm 0.0001$. To the best of our knowledge, this exceeds the score achieved by all previous demonstrations.\cite{CinelliPRL2005,YangPRL2005,BarbieriPRA2005,Aolita:2012PRA,Xu_PRL_2022,Sheffer2022} 
The uncertainty in each of the nine distinct games comes from a binomial distribution and is proportional to $1/\sqrt{N_s}$, and can be made much smaller by increasing $N_s$. The measured $P_w$ values exceed the classical bound owing to the non-commutativity of quantum observables. These scores show why quantum theory requires matrix representations, unlike classical physics with scalar observables. The threshold where quantum advantage vanishes marks where matrix mechanics is no longer needed.

\vspace{1mm}

\subhead{Quantifying contextuality}
A key feature of contextuality is that it enables state-independent tests. We therefore quantify contextuality in our quantum processors by an inequality, analogous to the Bell inequality. The Bell-Kochen-Specker (BKS) theorem states that no non-contextual hidden-variables (NCHV) theory can reproduce the predictions of quantum mechanics for correlations between measurement outcomes of some sets of observables\,\cite{Specker1960,Bell1966,KS1967,Mermin1993}. Cabello\,\cite{cabello2008experimentally} and others\,\cite{Badziabg_PRL_2009} derive a set of inequalities that are satisfied by any NCHV theory but are violated by quantum mechanics for any quantum state. These inequalities bound certain linear combinations of ensemble averages of correlations between measurement outcomes of compatible observables; thus creating a separation between the predicted outcomes of quantum mechanics, and the bound that is satisfied by NCHV theories. Consider the set of observables shown in Table 1, where observables within each column and each row commute. For any NCHV theory,

\begin{equation}
\label{eq:chi}
\chi_{\text{BKS}}= \langle R_1\rangle+ \langle R_2\rangle+\langle R_3\rangle-\langle C_1\rangle-\langle C_2\rangle-\langle C_3\rangle \le 4\,  
\end{equation}

\noindent where $\langle R_1\rangle$ is the ensemble average of the products of the outcomes of the observables listed in the first row, and so forth. The above inequality is independent of the preparation of the ensemble, provided all terms are estimated for the same preparation. 

\begin{figure*}[t!!]
\centering
\includegraphics[width=0.63\textwidth]{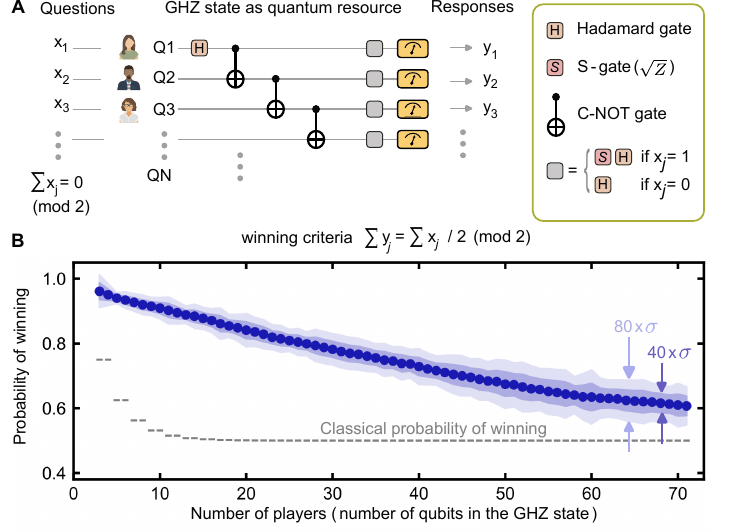}
\caption{\small{\textbf{The $N$-qubit GHZ parity game.} \textbf{(A)}, Quantum circuit for the game. Each of the $N$ players is given a binary value $x_j$ and is asked to return a binary answer $y_j$ such that the sum of the answers is the same,  { up to modulo 2}, as the sum of the questions divided by two. Using an $N$-qubit GHZ state, players can answer the questions by measuring in a corresponding basis, depending on the binary question ($x_j$) they were assigned. \textbf{(B)}, The average winning probability measured on up to 71 qubits (or players). 20 distinct games (a random set of $x_j$ values) were selected, and each game was implemented on 50 choices of GHZ state growth pattern. Each of these 1000 configurations was measured $N_s=1000$ times and returned a win or loss. The shaded regions are $\pm 20 \sigma$  and $\pm 40 \sigma$ of statistical error computed from the 50 qubit configurations used, after averaging over 20 games played, at each point\,(see Fig.~S6\,\cite{SM} for details). }}   
\end{figure*}

\vspace{1mm}
In Fig.\,2A we show a gate sequence that performs quantum non-demolition (QND) measurements of the operators in Table 1. Each row and column represents a distinct measurement context, and entails three successive measurements, which in theory should multiply to $+1$ for each row and $-1$ for each column. We run a circuit with a random sequence of 180 contexts and repeat this experiment 1000 times to gather statistics. The three measurements within each context are also performed in random order. This procedure is equivalent to evaluating the inequality over the maximally mixed state. In Fig.\,2B, we show the measured expectation value for each context (row or column, $6\langle R_i\rangle$ or $-6\langle C_i\rangle$). The sum of these averages gives $\chi_{\text{BKS}}=5.590\pm 0.008$, closer to the quantum limit of $\chi_{\text{BKS}}=6$ and exceeding previous experiments\cite{kirchmair2009state}, and exceeding the classical limit of $\chi_{\text{BKS}}=4$ by nearly 200 statistical standard deviations. Our result can also be expressed as a contextual fraction, $(5.59 -4) / 2 =  0.795$, where in ideal noiseless conditions, this fraction would be equal to 1.
The deviation from $6$ is caused by various experimental errors in our system\,(see\,\cite{SM} for details). 

It is worth emphasizing the connection between the two experiments presented thus far and comparing their outcomes. In general, any state-independent proof of strong local contextuality can be reformulated as a state-dependent nonlocal game\,\cite{Stairs:1983PS, heywood1983, aravind2002,Brown:1990FPH,abramsky2017quantum,cabello2021converting}. Although our implementations of these two approaches on noisy hardware are affected by different types of errors, both are well captured by our error-budget analysis (Figs.~S2 and S3). Thus, the observed success probabilities primarily reflect the distinct experimental ingredients and relative demands of each protocol.

\vspace{1mm}
\subhead{GHZ parity game}
To examine contextuality beyond few-body quantum states, we implement a non-communication game that can be won against classical strategies if the players share an $N$-qubit Greenberger-Horne-Zeilinger\,(GHZ) state\cite{GHZ1989} before playing the game. This scalable game offers a natural starting point for extending the discussion to many-body condensed matter systems\,\cite{Bulchandani_prb_2023_1,Bulchandani_PRB_2023_2}. In particular, it raises the question of whether such games can serve to verify contextuality in the ground states of physical systems. In the $N$-qubit GHZ parity game, there are $N$ players and each player $j$ is given a classical bit $x_j \in \{0,1\}$, the ``question,'' with the promise that $\sum_j x_j$ is even\,(Fig.\,3A). In order to win the game, the players must output bits $y_j \in \{0,1\}$ such that 

\begin{equation}
\sum_{j=1}^N y_j \overset{\textrm{mod 2}}{=} \frac{\sum_{j=1}^N x_j}{2}.
\end{equation}

\noindent If the players do not take advantage of quantum physics, the optimal strategy available to them is random guessing and wins with a probability $1/2 + 1/ 2^{\lceil N/2 \rceil}$ (gray line in Fig.\,3B, ref.\,\cite{brassard2005quantum}). If the players instead share the quantum state $\lvert \mathrm{GHZ} \rangle$ before playing the game, they can win with probability $P_w = 1$ by measuring their qubits in either the $X$ or $Y$ basis, depending on whether their input $x_j$ is 0 or 1. The players perform the following operations: { each player $j$} applies an $S$ gate to their qubit if $x_j = 1$; otherwise, no gate is applied. {Each player then applies a} Hadamard gate\,(gray squares in the quantum circuit), and measure their qubit in the $Z$ basis, which gives $y_j \in \{0,1\}$. Because the GHZ state, $ \lvert \mathrm{GHZ} \rangle = (\lvert 000... \rangle + \lvert 111... \rangle)/\sqrt{2}$, is stabilized by a tensor product of Pauli $X$ and $Y$ with an even number of $Y$, the results will always have even parity.

In Fig.\,3B, we show the results of implementing this algorithm on our superconducting processor up to $N=71$ players, where each player corresponds to a qubit. For each choice of the number of players, 20 distinct games (a given set of $x_j$) and 50 choices of qubits for each game are played, where each of these 1000 games is played $N_s=1000$ times, and each instance results in a definite win or loss. The winning probability is the average over all games with the indicated standard deviation coming from statistical error in each game. Although the measured probability of winning always stays above the classical winning probability, the margin of advantage narrows with the number of players. Nevertheless, if errors do not systematically favor the $ \left(\lvert 000... \rangle - \lvert 111... \rangle\right)/\sqrt 2$ state, playing each game more times reduces the standard deviation of the mean and increases the confidence in outperforming the classical strategy. { Our demonstration exceeds previous studies in scale and fidelity of GHZ state preparation \cite{Daniel2022, Hart_2025, Song2019, Gong2019,Mooney27_2021,Mooney65_2021}. Notably, our genuine entanglement reaches 45 qubits (see Fig.~S6\,\cite{SM}), which is larger than the full system size of these demonstrations. Other studies with more than 45 and up to 65 qubits \cite{Bao2024, Cao2023, Mooney65_2021} have also been reported. The only experiment with a larger qubit count than ours is that reported in  \cite{javadi120_2025}, where postselection and other error mitigation techniques, incompatible with the incommunicado GHZ game that interest us here, were used.}

\begin{figure*}[t!!]
\centering
\includegraphics[width=0.98\textwidth]{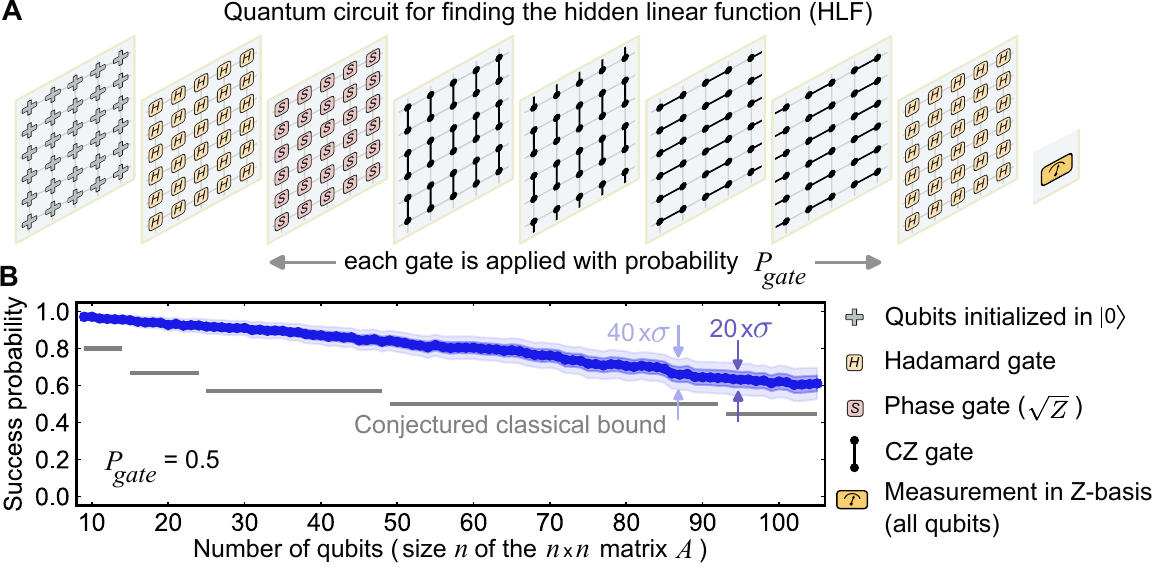}
\caption{\small{{\textbf{Quantum solution to the 2D  HLF problem.}} \textbf{(A)}, Quantum circuit for finding a hidden linear  function. Qubits are initialized in a product $|0\rangle$ state and then a Hadamard
gate is applied to all of them. The encoding of the symmetric $n\times n$ matrix $A$ is done by applying an $S$-gate where $A_{i,i}$ is non-zero and CZ-gate between qubits $i$ and $j$ where $A_{i,j}$ is nonzero. The diagonal and allowed off-diagonal elements in $A$ are non-zero with probability 0.5, where the allowed off-diagonal elements are those which connect neighboring qubits. At the end, all qubits are measured in the computational basis, after a layer of Hadamard gates on all qubits.\,\textbf{(B)}, The {probability of success in finding the solution for the HLF problem with our quantum computer} plotted for various $n$ values up to 105. In the absence of errors, the algorithm returns a correct hidden bitstring with probability one. Data points are the average of $N_g=1000$ distinct matrices, each run $N_s=100$ times, at each $n$-value, and the standard deviation $\sigma$ is the statistical fluctuations of  $N_g$ distinct runs\,(see Fig.~S7\,\cite{SM} for details). The gray horizontal lines {mark the probability threshold where we believe quantum advantage over a classical approach is lost, conjectured from the assumption that the hidden linear function depends on { every} input { and that solving it requires classical circuits of depth scaling with $\log_{2}(s)$ (see the main text)}.}}} 
\label{fig:shallow_circuits}
\end{figure*}

\subhead{Hidden linear function problem}
{ We next pivot to a large-scale experiment where contextuality bounds have not yet been discovered. In 2018, an unconditional quantum advantage was predicted for a restricted class of circuits by Bravyi et al. \cite{bravyi2018quantum}, and it has so far remained experimentally untested --see, however, a pioneering 6-qubit demonstration reported in \cite{Daniel2022}--.  Bravyi et al.'s proof is only valid for tens of thousands of qubits and, therefore, an unconditional demonstration of quantum advantage will remain inaccessible for the foreseeable future. Here we report on a large-scale experimental test of the algorithm.} 

The computational problem that Bravyi et al.~considered can be viewed as a non-oracular variant of the well-known Bernstein-Vazirani problem\,\cite{bernstein1997quantum} {where one is tasked to find a solution to a hidden linear function (HLF) problem}. Formally, consider quadratic forms $q: \{0,1\}^n \rightarrow \mathbb{Z}_4 = \{0,1,2,3\}$, defined as 

\begin{equation}
q(x) \overset{\textrm{mod 4}}{=}
 \sum_{i,j=1}^n A_{i,j} x_i x_j,
\end{equation}

\vspace{1mm}
\noindent where $x = (x_1, \ldots, x_n)$ is a vector of $n$ binary variables and $A$ is a symmetric binary matrix. We consider the following HLF problem: Given an $n{\times}n$ symmetric binary matrix $A$ specifying a quadratic form $q(x) = x^T A x$, find a binary vector 
$z \in \{0,1\}^n$ such that $q(x) = 2 z^T x$ (mod 4)  for all $x$ in the binary null-space of $A$. Bravyi et al.~devised a quantum algorithm for solving this 2D problem.\,\cite{bravyi2018quantum} They assign each input bit $A_{i, j}$ to the edge $\{{i, j\}}$ and each bit $A_{i, i}$ at the vertex $i$ of the 2D grid of qubits. They showed that the 2D HLF problems that are mappable to a 2D square grid of qubits with local gates can be solved with certainty by a shallow quantum circuit. { As argued by the authors and others \cite{bravyi2018quantum,aasnass2022,Vieiraetal2025}, the source of the advantage can be tracked back to quantum contextuality.}

In Fig.~4A, we show the pulse sequence used for solving a 2D HLF problem for an $n {\times} n$ symmetric binary matrix. Qubits are initialized in the $\lvert 0\rangle$  state and then a Hadamard gate is applied to all qubits to create a superposition of all computational states. We focus on a subset of random binary $n\times n$ matrices $A$ in which each connectivity-allowed entry is non-zero with probability $P_{\rm gate}$. To implement the circuit associated with such a matrix, we apply S gates followed by four layers of CZ gates, where S and CZ gates are included whenever the corresponding elements of $A$ are non-zero. Next, there is a layer of Hadamard gates, followed by a measurement of all qubits, where each measurement instance returns a bitstring that one can verify by comparing against the classical simulation of the quantum circuit. 

{The four-layer circuit of Bravyi et al. is affected by noise, resulting in a reduced success probability on the HLF problem. In Fig.~4B, we show the probability of success for our noisy quantum circuit resulting from measuring $N_g=1000$ {random} choices of $A$ and measuring each choice $N_s=100$ times for matrix size $n$ values ranging from 9 to 105 qubits. Running the four layers circuit until success provides a solution, which can be efficiently classically verified using Stim\,\cite{gidney2021stim}.} 

{If extrapolated down to only a few hundred qubits, Bravyi et al.'s classical bound results are not informative as they suggest that any classical circuit solving the problem with high probability should have depth $\geq1$, which is a trivial statement. Therefore, in the absence of a theoretical provable lower bound for our experiment, we propose an intuitive measure below.}
 {We conjecture that { the minimum classical circuit depth required to implement a Boolean function, using two-input--one-output gates, that depends on all $s$ inputs and solves the HLF problem, is $\log_{2}(s)$.} The formula is obtained by counting the number of two-bit gate layers a classical computer requires to find a length-n solution bitstring, and does not limit the connectivity of classical circuits. We further highlight that the comparison is slightly disadvantageous to the quantum computer in the sense that quantum computer connectivity is limited, whereas classical circuit connectivity is not. In Fig.~4B we plot with horizontal lines the qubit-dependent threshold where one would need to run, on average, more than $\log_{2}(s)$ quantum layers. In this sense, the problem is effectively solved using fewer two-qubit gate layers on our quantum processor than two-bit gate layers in the classical circuit model. We acknowledge that there is an opportunity here to theoretically derive formal bounds for our current (and foreseeable future) experiments, in order to evaluate to what degree noise hampers the possible quantum advantage.}

\subhead{Discussion and outlook}
In this work, we leveraged the flexibility of a large 2D array of superconducting qubits to show that many-body quantum contextuality can be used to perform bounded-resource tasks { more efficiently than what can be achieved classically under the working hypothesis}. At the few-qubit level, we implemented the magic square game and violated a Bell–Kochen–Specker (or non-contextuality) inequality. In this regime, we achieved performance that improves upon prior work\,\cite{kirchmair2009state,CinelliPRL2005,YangPRL2005,BarbieriPRA2005,Aolita:2012PRA,Xu_PRL_2022,Sheffer2022}, where the maximum attainable contextuality is now the target rather than merely the violation of a lower bound. We extended these demonstrations to scalable many-body tasks, advancing the state of the art in both fidelity and number of qubits, and demonstrated contextual behavior using the $N$-qubit GHZ parity game to outperform the classical strategy with up to 71 qubits. { We also provide the data for a {large-scale} demonstration for the implementation of the HLF problem, with success probability higher than 0.5 up to 105 qubits. Our experiment implements, on a large scale, a quantum algorithm designed to find an HLF; it represents a substantial advancement over a previous demonstration \cite{Daniel2022}.} The sensitivity of these many-body algorithms also allowed us to benchmark our processor and detect subtle coherent errors that do not appear in random unitary gate operations. To further inspire hardware development, i.e., to reduce the performance gap from the noiseless quantum expectation, { we propose using the quantitative many-body contextuality benchmarks we provide here for current and future quantum devices.}

\newpage
\onecolumngrid

\begin{flushleft}
{\hypertarget{authorlist}{${}^\dagger$}  \small Google Quantum AI and Collaborators}

\bigskip

    \renewcommand{\author}[2]{#1\textsuperscript{\textrm{\scriptsize #2}}}
    \renewcommand{\affiliation}[2]{\textsuperscript{\textrm{\scriptsize #1} #2} \\}
    \newcommand{\corrauthora}[2]{#1$^{\textrm{\scriptsize #2}, \hyperlink{corra}{\ddagger}}$}
    \newcommand{\corrauthorb}[2]{#1$^{\textrm{\scriptsize #2}, \hyperlink{corrb}{\mathsection}}$}

\begin{footnotesize}

\newcommand{\xGoogle}{\affiliation{1}{Google Research, Mountain View, CA, USA}}

\newcommand{\xPrincetonEE}{\affiliation{2}{Department of Electrical and Computer Engineering,
Princeton University, Princeton, NJ, USA}}

\newcommand{\xYale}{\affiliation{3}{Department of Applied Physics, Yale University, New Haven, CT, USA}}

\newcommand{\xUMass}{\affiliation{4}{Department of Electrical and Computer Engineering, University of Massachusetts, Amherst, MA}}

\newcommand{\xSevilla}{\affiliation{5}{Instituto Carlos I de F\'{\i}sica Te\'orica y Computacional, Universidad de Sevilla, E-41012 Sevilla, Spain}}

\newcommand{\xSevillaII}{\affiliation{6}{Departamento de F\'{\i}sica Aplicada II, Universidad de Sevilla, E-41012 Sevilla, Spain}}

\newcommand{\xMinnesota}{\affiliation{7}{School of Physics and Astronomy, University of Minnesota, Minneapolis, Minnesota 55455, USA}}

\newcommand{\xRice}{\affiliation{8}{Department of Physics and Astronomy, Rice University, 6100 Main Street Houston, TX 77005, USA}}

\newcommand{\xOxford}{\affiliation{9}{Rudolf Peierls Centre for Theoretical Physics, University of Oxford, Oxford OX1 3PU, United Kingdom}}

\newcommand{\xUCSB}{\affiliation{10}{Department of Physics, University of California, Santa Barbara, CA}}

\newcommand{\Google}{1}
\newcommand{\PrincetonEE}{2}
\newcommand{\Yale}{3}
\newcommand{\UMass}{4}
\newcommand{\Sevilla}{5}
\newcommand{\SevillaII}{6}
\newcommand{\Minnesota}{7}
\newcommand{\Rice}{8}
\newcommand{\Oxford}{9}
\newcommand{\UCSB}{10}

\corrauthora{S. ~Kumar}{\Google, \!\PrincetonEE},
\corrauthora{E. Rosenberg}{\Google},
\corrauthora{A. Grajales~Dau}{\Google},
\corrauthora{R. G.~Cortiñas}{\Google},
\author{D. Maslov}{\Google},
\author{R. Oliver}{\Google}, 
\author{A. Zalcman}{\Google},
\author{M. Neeley}{\Google},
\author{A. Pagano}{\Google},
\author{A. Szasz}{\Google},
\author{I. Drozdov}{\Google},
\author{Z. Minev}{\Google},
\author{C. Gidney}{\Google},
\author{N. Yosri}{\Google},
% ...
\corrauthorb{P. Roushan}{\Google},
\corrauthorb{M. Devoret}{\Google,\!\UCSB}

\bigskip

\xGoogle
\xPrincetonEE
\xYale
\xUMass
\xSevilla
\xSevillaII
\xMinnesota
\xRice
\xOxford
\xUCSB

\vspace{2mm}
{\hypertarget{corra}{${}^\ddagger$} These authors contributed equally to this work.}\\

{\hypertarget{corrb}{${}^\mathsection$} Corresponding author: pedramr@google.com}\\
{\hypertarget{corrb}{${}^\mathsection$} Corresponding author: devoret@google.com}

\end{footnotesize}
\end{flushleft}

\twocolumngrid
\bibliography{References.bib}

\onecolumngrid
\newpage

\vspace{3mm}
\noindent\textbf{Acknowledgments.} We thank Č. Brukner, B. Dakić, C. A. Fuchs, R. M. Nandkishore, N. Medina Sanchez, and R.~Spekkens for insightful discussions.

\vspace{1mm}
\noindent\textbf{Funding:} A.C. acknowledges support from the EU-funded project FoQaCiA. S.L.S. was supported by a Leverhulme International Professorship (LIP-202-014) and by EPSRC via grant EP/X030881/1.

\vspace{1mm}
\noindent\textbf{Author contributions:} M.D. and P.R. proposed the experiment, guided the research, and helped interpret the results. S.K., E.R., A.G.D., and R.G.C. contributed equally to this work by designing and executing the experiments, calibrating the device, and analyzing the data. A.C., S.L.S., V.B.B., and F.J.B. provided theoretical support. All authors contributed to writing the manuscript. The Google Quantum AI team provided the experimental infrastructure that enabled this work.

\vspace{1mm}
\noindent\textbf{Competing interests:} The authors declare no competing interests.

\vspace{1mm}
\noindent\textbf{Data and materials availability:} All data shown in this work are available on Zenodo\,\cite{data_zenodo}. The code used to generate and analyze the data is available as part of ReCirq\,\cite{recirq}. No new materials were generated for this study.

\smallskip

\textbf{Supplementary materials}
\begin{itemize}
    \item[] Materials and Methods
    \item[] Supplementary Text 
    \item[] Figs.~S1-S7
    \item[] References (85--91)
\end{itemize}

\nocite{google_y1,optimizing_klimov_2024,model-based_optimization_bengtsson_2024,cirq_zenodo_2024,GHZ_kandala,GHZ_flammia2011direct,BFS_cormen2022}

\clearpage
\onecolumngrid
\setcounter{section}{0}\setcounter{subsection}{0}\setcounter{figure}{0}\setcounter{equation}{0}\setcounter{table}{0}
\makeatletter
\renewcommand{\thesection}{\arabic{section}}
\renewcommand{\thesubsection}{\thesection.\Alph{subsection}}
\renewcommand{\thesubsubsection}{\alph{subsubsection}}
\renewcommand{\thefigure}{S\@arabic\c@figure}
\renewcommand{\theequation}{S\@arabic\c@equation}
\renewcommand{\thetable}{S\@arabic\c@table}
\renewcommand{\theHsection}{S\arabic{section}}
\renewcommand{\theHfigure}{S\arabic{figure}}
\renewcommand{\theHequation}{S\arabic{equation}}
\renewcommand{\theHtable}{S\arabic{table}}
\makeatother
\begin{center}
{\large\bfseries Supplementary Materials for\\``Quantum--Classical Separation in Bounded-Resource Tasks Arising from Measurement Contextuality''}\\[2ex]
Google Quantum AI and Collaborators
\end{center}

\vspace{1em}
\noindent{\large\bf This PDF file includes:}

\smallskip
\noindent Materials and Methods\\
Supplementary Text\\
Figs.~S1 to S7\\
References (85--91)

\newpage

\section{Materials and Methods}
\label{sec:materials_methods}

\subsection{Device characterization}
\label{sec:exp_tech}

\begin{figure*}[b!]
\centering
\includegraphics{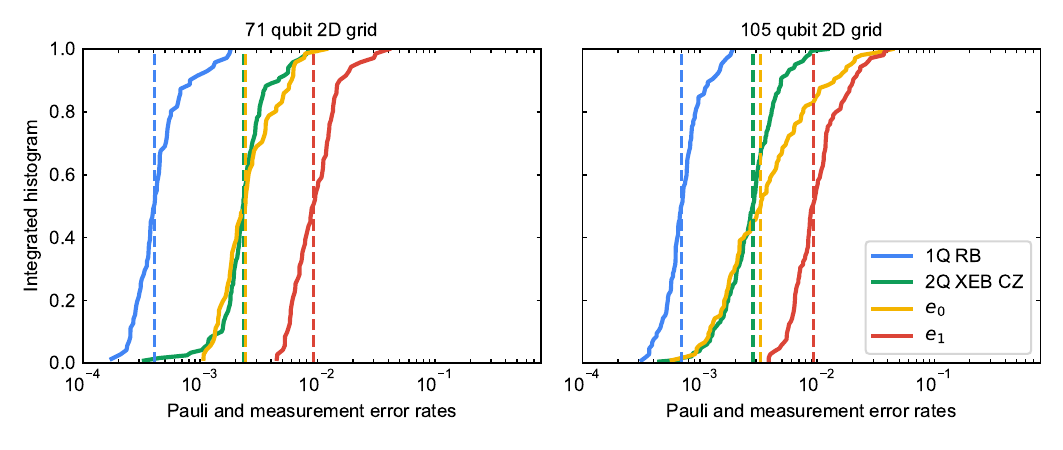}
\caption{Representative error rates for a 71-qubit 2D grid on the 72-qubit Willow processor and a 105-qubit 2D grid on the 105-qubit Willow processor. The data include readout (measurement) errors, single-qubit randomized benchmarking (1Q RB) errors, and two-qubit cross-entropy benchmarking (2Q XEB) Pauli errors for CZ gates. Vertical dashed lines denote the median values for each dataset. The readout errors $\mathbf{e_0}$ and $\mathbf{e_1}$ correspond to the probability of reporting $|1\rangle$ when the qubit is prepared in $|0\rangle$ and of reporting $|0\rangle$ when prepared in $|1\rangle$, respectively.
}
\label{sfig:error_rates}
\end{figure*}

The experiments were performed on 72-qubit and 105-qubit Willow processors, with architectures similar to those described in Ref.\,\cite{google_y1}. In all cases, CZ gates served as the native entangling operation. Fig.~\ref{sfig:error_rates} shows representative error rates for a 71-qubit grid on the 72-qubit processor and a 105-qubit grid on the 105-qubit processor. Details of gate optimization are provided in Refs.\,\cite{optimizing_klimov_2024, model-based_optimization_bengtsson_2024}.

\newpage

\section{List of symbols}
\label{sec:symbols}
\renewcommand{\arraystretch}{1.5}
\begin{center}
\begin{tabular}{ |l|l| } 
 \hline
 Symbol & Description\\ 
 \hline
$N$ & Number of qubits in the GHZ state \\
$\ket{\mathrm{GHZ}}$ & $N$-qubit GHZ state, $\tfrac{1}{\sqrt{2}}\big(\ket{0}^{\otimes N}+\ket{1}^{\otimes N}\big)$ \\
$\tilde{\rho}_{\mathrm{GHZ}}$ & Experimentally prepared GHZ state \\
$F$ & Fidelity of $\tilde{\rho}_{\mathrm{GHZ}}$ with respect to $\ket{\mathrm{GHZ}}$ \\
$\hat P_i$ & $N$-qubit Pauli operator \\
$\{\hat S_i\}$ & Set of $2^N$ stabilizers of $\ket{\mathrm{GHZ}}$ \\
$\hat Q_i$ & Generator of the stabilizer group \\
$F^X, F^Z$ & Fidelities associated with $X$-type and $Z$-type stabilizers \\
$F_{T_1}$ & Fidelity of GHZ state under $T_1$ decay \\
$n_{\mathrm{sq}}$ & Number of single-qubit gates \\
$e_p^{\mathrm{sq}}$ & Pauli error probability per single-qubit gate \\
$F_{\mathrm{sq}}, F_{\mathrm{sq}}^{X}, F_{\mathrm{sq}}^{Z}$ & Total, $X$-type, and $Z$-type fidelities from single-qubit errors \\
$n_{\textrm{2q}}$ & Number of two-qubit gates \\
$e_p^{\mathrm{2q}}$ & Pauli error probability per two-qubit gate \\
$F_{\mathrm{2q}}, F_{\mathrm{2q}}^{X}, F_{\mathrm{2q}}^{Z}$ & Fidelities from two-qubit gate errors \\ $n_{\mathrm{ro}}$ & Number of measurement gates \\
$e_0, e_1$ & Probabilities of misidentifying $\ket{0},\ket{1}$ in readout \\
$\varepsilon$ & Symmetric readout error rate ($e_0=e_1=\varepsilon$) \\
$Z_\Lambda, X_\Lambda, Y_\Lambda$ & Transformed observables under readout noise \\
$F_{\mathrm{readout,asym}}$ & Fidelity under asymmetric readout error \\
$F_{\mathrm{readout,sym}}, F_{\mathrm{readout,sym}}^{X}, F_{\mathrm{readout,sym}}^{Z}$ & Fidelities under symmetric readout error \\
$F_{\mathrm{total}}$ & Total GHZ fidelity including all error channels \\
$P_{\mathrm{w}}$ & Winning probability of the bounded-resource task\\
$N_{g}$ & Number of randomized GHZ game or HLF instances \\
$N_{s}$ & Number of shots (repetitions) per instance \\
$SD_{50},SD_{1000}$ & Standard deviation across 50 GHZ or 1000 HLF instances \\
$L_{\mathrm{classical}}$ & Lower bound on classical circuit depth \\
$A$ & $n \times n$ matrix defining the 2D HLF problem \\
$S$ & Phase gate ($\sqrt{Z}$) used in the HLF encoding \\
CZ & Controlled-$Z$ gate (native two-qubit entangling operation) \\
$P_{\rm gate}$ & Probability of applying each allowed $S$ or CZ gate in the 2D HLF problem \\
\hline
\end{tabular}
\end{center}

\newpage

\section{Magic Square Game: Variations and Error Budget}
\label{sec:mermin_peres_supp_material}

\begin{figure*}[b!]
\centering
\includegraphics{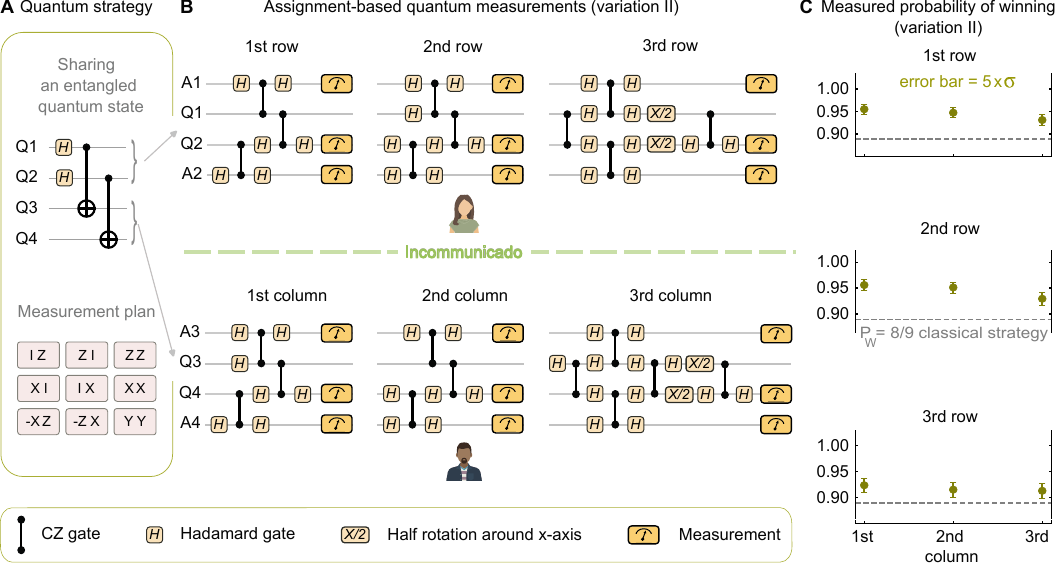}
\caption{\textbf{Variation II of the magic square game.} \textbf{(A)} Players share two Bell pairs and agree on the measurement basis for each assigned row (Alice) or column (Bob). \textbf{(B)} Using ancilla qubits, each player performs three simultaneous measurements corresponding to their assigned row or column. For clarity, we omit the dynamical decoupling sequences applied to qubits during idle periods. We also depict the circuits using CZ, Hadamard, and X/2 gates, although in the actual implementation, contiguous single-qubit gates are compiled into a single operation.
 \textbf{(C)} Measured probability that both players follow their respective multiplication rules and obtain matching outcomes in the shared cell, shown for nine distinct games (row-column assignments) sorted by the position of their intersection. The average win probability is $P_{\mathrm{w}} = 0.9355 \pm 0.0008$. Each game is played $N_s = 10{,}000$ times, and the error bar for each game is five times the statistical uncertainty derived from the binomial distribution.}
\label{sfig:mp_game_scheme_ii}
\end{figure*}

When using quantum resources, the magic square game can be played in several different ways that would all be equivalent in the absence of noise:

\begin{itemize}
 \item \textbf{Variation I.} Alice and Bob decide to measure only two of the three Pauli operators corresponding to their assigned row and column. They then each infer the outcome of the third Pauli such that the product of the three results equals $+1$ for Alice (assigned to rows) and $-1$ for Bob (assigned to columns). This is the strategy that is used in Figure 1 of the main text.

 \item \textbf{Variation II.} In this version, they do not compute the third Pauli outcome but measure it directly on the processor (Fig.~\ref{sfig:mp_game_scheme_ii}). Consequently, the circuits become larger to accommodate the third Pauli measurement. Each player uses two ancilla qubits to measure two Paulis and employs their half of the Bell pair to measure the third Pauli. This circuit construction is not unique and could alternatively be done with one ancilla. Either the rules of the game or the strategy utilized by Alice and Bob must be modified to accommodate cases where the three measurements do not multiply correctly (which was stipulated as a rule in the original formulation of the game), but here we focus on presenting the experimental data itself rather than discussing the implications of variations of the game's rules. 
\end{itemize}

\begin{figure*}[b!]
\centering
\includegraphics{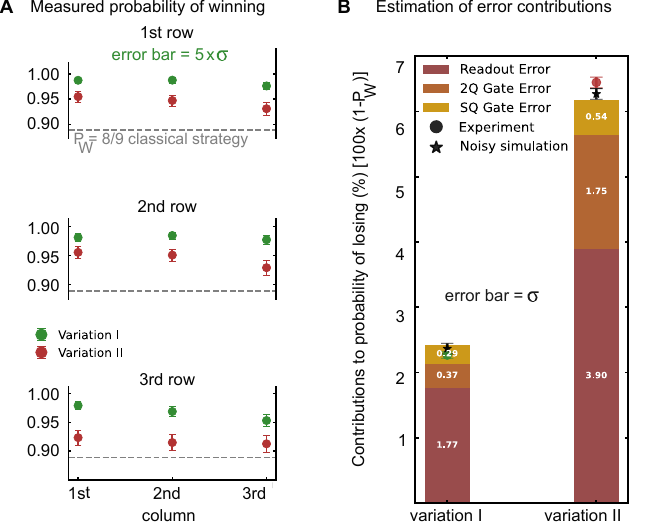}
\caption{\textbf{Comparing the performance of game variations.} \textbf{(A)} Win probabilities for Variation I (green) and Variation II (red) across the 9 distinct games ordered by the corresponding row-column intersection. The average win probability is $P_w = 0.9773 \pm 0.0005$ for Variation I and $P_w = 0.9355 \pm 0.0008$ for Variation II. Each game is played $N_s = 10{,}000$ times. The error bars represent five times the standard error. \textbf{(B)} Colored bars represent the simulated contributions of individual error mechanisms to the total probability of losing. Black stars show simulations with all mechanisms enabled. Points (green/red) show the measured average loss probability for each variation. Error bars in this panel indicate the standard error.}
\label{sfig:mermin_peres_error_budget}
\end{figure*}

\newpage
\noindent Figure~\ref{sfig:mermin_peres_error_budget}(A) shows the measured win probabilities for Variations I and II. Panel (B) presents the estimated error budget for both variations, in agreement with the average measured probability of winning (green/red points). We model each mechanism separately (colored bars) by either adding depolarizing noise following the single-qubit (SQ) or two-qubit (2Q) gates, using Cirq\,\cite{cirq_zenodo_2024}, or by randomly flipping bits in the simulated bit strings to model readout errors. These error processes are simulated with the measured single-qubit Pauli error $e_p^{\mathrm{sq}}$, two-qubit Pauli error $e_p^{\mathrm{2q}}$, and readout error rates $e_0$ ($0\to1$) and $e_1$ ($1\to0$) for the qubits involved. We also perform simulations with all error mechanisms enabled simultaneously (black stars), which are consistent with the per-mechanism estimates within uncertainties. Because Variation~II directly measures the third Pauli in each context, it requires additional gates and therefore exhibits a higher loss rate. Simulations are performed with 10,000 shots and are therefore subject to statistical uncertainties, which are reflected in the error bars on the black stars. These error bars do not reflect uncertainty due to imprecision in measurements of the physical error rates or due to limitations of the model (e.g. errors may not be exactly depolarizing, and other sources of error such as $T_1$ decay were not included). Nevertheless, the agreement with the experimental win rate suggests that we have correctly captured the major sources of error.

\newpage

\newpage

\section{Bell-Kochen-Specker Inequality: Further Data and Calibration}
\label{sec:KSB}

\subsection{Continuous Measurement of a Random Sequence}
\label{ssec:further_data_KSB}

To evaluate \(\chi_{\mathrm{BKS}}\), we stream a random sequence of 180 measurement contexts, rows or columns from the Mermin-Peres square (Table 1 of the main text), and collect statistics over $N_s=1000$ runs of such sequences. We note that this procedure is equivalent to evaluating $\chi_{\rm BKS}$ on the maximally mixed state. After measuring any row or column of the magic square, the system is projected onto an eigenstate of the corresponding commuting operators. Because all such eigenstates occur with equal probability, averaging over many experimental runs yields an effective state that is exactly maximally mixed.

\begin{figure*}[b!]
 \centering
 \includegraphics{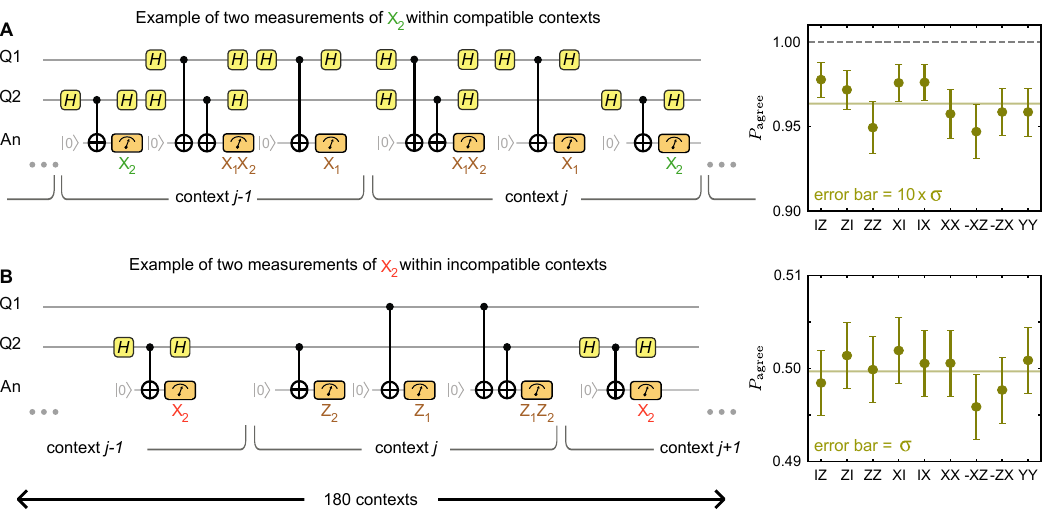}
 \caption{\textbf{(A)} Example segment where a Pauli (here \(X_2\)) is measured twice with only commuting operators between contexts \(j-1\) and \(j\), implying \(P_{\mathrm{agree}}=1\). Right: \(P_{\mathrm{agree}}\) for each Pauli over all such compatible occurrences. The horizontal solid line is the average $P_{\mathrm{agree}} = 0.9636 \pm 0.0004$. The error bars in the plot are \(10\times \sigma\) for visibility. \textbf{(B)} Example segment where intermediate operators do not commute with \(X_2\), implying \(P_{\mathrm{agree}}=1/2\). Right: \(P_{\mathrm{agree}}\) for each Pauli over all such incompatible occurrences. The horizontal solid line is the average $P_{\mathrm{agree}} = 0.500 \pm 0.001$.}
 \label{sfig:p_agree}
\end{figure*}

If a measurement Pauli $P$ appears in two consecutive compatible contexts, that is, with only commuting operators in between (compatible contexts, e.g., \(X_2\) in Fig.~\ref{sfig:p_agree}(A), green), the two outcomes, in principle, should agree, so \(P_{\mathrm{agree}}=1\). We experimentally measure \(P_{\mathrm{agree}}\) for every Pauli over all such compatible occurrences and plot the results in Fig.~\ref{sfig:p_agree}(A) (right). The binomial standard error is 

\begin{equation}
\sigma=\sqrt{\frac{P_{\mathrm{agree}}\bigl(1-P_{\mathrm{agree}}\bigr)}{N_c}},
\label{eq:ks_sigma}
\end{equation}
with \(N_c=20{,}000\), the average number of occurrences of each Pauli. If a given Pauli (e.g., \(X_2\) in Fig.~\ref{sfig:p_agree}(B), red) is measured before and after a context incompatible with it (\(Z_2\), \(Z_1\), \(Z_1 Z_2\)), the two measurement results should be fully uncorrelated. For such instances, the measurements for all Paulis are shown in Fig.~\ref{sfig:p_agree}(B) (right), with uncertainties from Eq.~\eqref{eq:ks_sigma}.

\vspace{20mm}

\newpage
\subsection{Calibrating Compensation Phases }

\begin{figure*}[b!]
\centering
\includegraphics{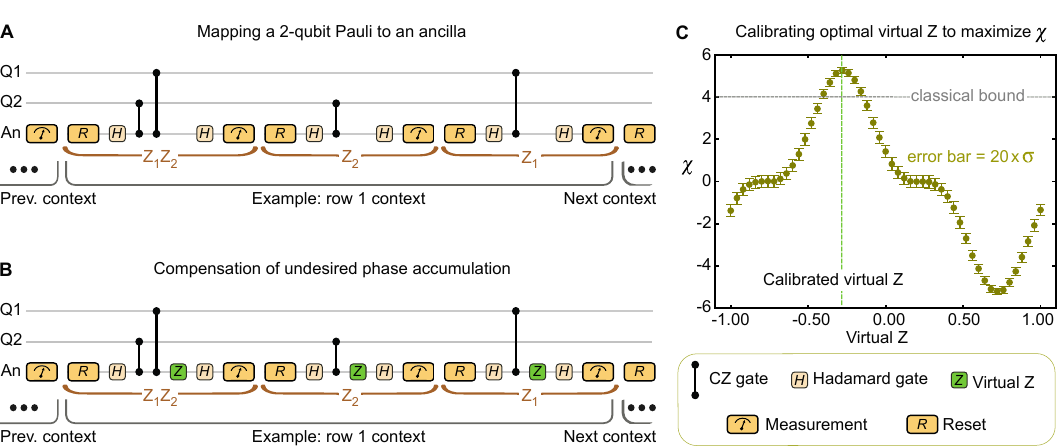}
\caption{\textbf{Compensating undesired phase accumulation.} \textbf{(A)} Example segment within the 180-context sequence where we measure row 1. \textbf{(B)} To compensate for the undesired phase accumulation, we introduce virtual Z gates before each final Hadamard during a Pauli mapping. \textbf{(C)} Calibrating the optimal virtual Z\,(\,applied in \textbf{(B)}\,) that will maximize the value of $\chi_{\mathrm{BKS}}$. The error bars plotted are 20 times the statistical uncertainty derived from the binomial distribution.}
\label{sfig:virtual_z_circuit}
\end{figure*}

To continuously measure the two-qubit Pauli operators that enter $\chi_{\mathrm{BKS}}$, we map the joint state of Q1 and Q2 onto the phase of an ancilla qubit. Figure~\ref{sfig:virtual_z_circuit}(A) illustrates the circuit used to implement this mapping. In this construction, the ancilla is highly sensitive to undesired phase accumulation. Between successive two-qubit Pauli measurements, the ancilla is measured and reset, a process that requires large frequency excursions in our architecture. The finite settling time associated with these excursions introduces additional phase accumulation on the qubits, which in turn produces a distribution of $\chi_{\mathrm{BKS}}$ values that depends on the flux-line transfer function of each ancilla qubit.

To compensate for this effect, we insert a virtual $Z$ rotation before the final Hadamard in each two-qubit Pauli-mapping sequence(Fig.~\ref{sfig:virtual_z_circuit}(B)) and calibrate its value to cancel the accumulated phase. Figure~\ref{sfig:virtual_z_circuit}(C) shows the estimated $\chi_{\mathrm{BKS}}$ as a function of the applied virtual $Z$.

\newpage

\newpage

\newpage

\section{GHZ State: Fidelity and Error Estimation}
\label{sec:ghz_stab_error_budget}

\subsection{Fidelity}
\label{ssec:ghz_fidelity}

\subsubsection{Definition}
\label{sssec:ghz_fid_def}

We aim to estimate the fidelity of the experimentally prepared state, $\tilde{\rho}_{\mathrm{GHZ}}$, with respect to the $N$-qubit GHZ state, 
\begin{equation}
\ket{\mathrm{GHZ}} = \frac{ \ket{0}^{\otimes N} + \ket{1}^{\otimes N}}{\sqrt{2}}.
\label{eq:ghz_def}
\end{equation}
\noindent The fidelity is defined as
\begin{equation}
F \equiv \bra{\mathrm{GHZ}} \tilde{\rho}_{\mathrm{GHZ}} \ket{\mathrm{GHZ}}.
\label{eq:ghz_fidelity_def}
\end{equation}

\noindent Expanding the state $\tilde \rho_{\rm GHZ}$ in the Pauli basis,
\begin{equation}
\tilde{\rho}_{\mathrm{GHZ}} = \frac{1}{2^N} \mathbb{I} + \sum_{i=1}^{4^N-1} c_i \,\hat P_i,
\label{eq:pauli_expansion}
\end{equation}
where $\hat P_i$ are $N$-qubit Pauli operators, and the $c_i$ are the real-valued Pauli expansion coefficients, and substituting this expansion into Eq.~\eqref{eq:ghz_fidelity_def} gives
\begin{equation}
F = \frac{1}{2^N} + \sum_{i=1}^{4^N-1} c_i \,\bra{\mathrm{GHZ}} \hat P_i \ket{\mathrm{GHZ}}.
\label{eq:F_from_pauli}
\end{equation}

\noindent Since the GHZ state is a stabilizer state, it has $2^N$ stabilizer operators $\{\hat S_i\}$. For these operators
\begin{equation}
\bra{\mathrm{GHZ}} \hat S_i \ket{\mathrm{GHZ}} = 1,
\label{eq:stab_ev_one}
\end{equation}
whereas for all other Pauli operators $\hat P'_i$, 
\begin{equation}
\bra{\mathrm{GHZ}} \hat P'_i \ket{\mathrm{GHZ}} = 0.
\label{eq:nonstab_zero}
\end{equation}
The fidelity $F$ can therefore be expressed in terms of the stabilizer coefficients as
\begin{equation}
F = \sum_{i=0}^{2^N-1} c_i,
\label{eq:F_sum_cj}
\end{equation}
where the sum is restricted to the $2^N$ stabilizers. In particular, $c_0$ is the expansion coefficient of the identity stabilizer, $c_0 = 1/2^N$. From Eq.~\eqref{eq:pauli_expansion}, these coefficients are
\begin{equation}
c_i = \frac{1}{2^N} \Tr\!\left[\tilde{\rho}_{\mathrm{GHZ}} \,\hat S_i\right].
\label{eq:cj_from_trace}
\end{equation}
Thus, the fidelity is the arithmetic mean of the stabilizer expectation values
\begin{equation}
F = \frac{1}{2^N} \sum_{i=0}^{2^N-1} \langle \hat S_i \rangle,
\label{eq:final}
\end{equation}
where $\langle \hat S_i \rangle$ denotes the expectation value of stabilizer $\hat S_i$ in state $\tilde{\rho}_{\mathrm{GHZ}}$. As shown in Refs.\,\cite{GHZ_kandala,GHZ_flammia2011direct}, this quantity can be estimated by Monte Carlo sampling over a polynomial number of stabilizers, thereby avoiding the exponential cost of full $N$-qubit tomography.

\subsubsection{Stabilizers}
\label{sssec:ghz_stabilizers}

The $N$-qubit GHZ state $\ket{\mathrm{GHZ}}$ is a stabilizer state with $2^N$ stabilizers generated by $N$ independent stabilizers, which can be chosen to be
\begin{equation}
\hat Q_i =
\begin{cases}
\hat Z_i \hat Z_{i+1}, & i \in \{0,1,\ldots,N-2\}, \\[6pt]
\displaystyle \prod_{j=0}^{N-1} \hat X_j, & i = N-1.
\end{cases}
\label{eq:stabilizer_generators}
\end{equation}
The stabilizers $\hat S_i$ can be written in terms of these generators as 
\begin{equation}
\hat S_i = \hat Q_{N-1}^{\,i_{N-1}} \hat Q_{N-2}^{\,i_{N-2}} \cdots \hat Q_0^{\,i_0},
\label{eq:stabilizers}
\end{equation}
where $i_k$ denotes the $k$th binary digit of $i$. The $2^{N-1}$ stabilizers with $i_{N-1}=0$ are \emph{$Z$-type stabilizers}, diagonal in the computational basis. The remaining $2^{N-1}$ stabilizers with $i_{N-1}=1$ are \emph{$X$-type stabilizers}, which involve measurements in rotated bases $X$ and $Y$. The total fidelity can be expressed in terms of the average fidelities of these two classes as
\begin{equation}
F = \frac{F^X + F^Z}{2}.
\label{eq:F_sum_FX_FZ}
\end{equation}

\subsection{Error Estimation}
\label{ssec:ghz_error_estimation}

\subsubsection{Fidelity under $T_1$ with Dynamical Decoupling}
\label{sssec:ghz_fidelity_T1}

To estimate the 
fidelity of the GHZ state, we consider a mean relaxation time $T_1$ for all qubits. Let $\tau$ denote a small idle interval, the amplitude-damping rate is
\begin{equation}
\gamma \equiv\frac{\tau}{T_1}.
\label{eq:gamma}
\end{equation}

\noindent A general single-qubit density matrix, with unit trace and positivity constraints, takes the form
\begin{equation}
\rho = \begin{pmatrix} p_0 & c \\ c^\ast & p_1 \end{pmatrix},
\qquad p_0,p_1 \ge 0,\quad p_0 + p_1 = 1,\quad |c|^2 \le p_0 p_1.
\label{eq:rho_general}
\end{equation}
Here $p_0$ and $p_1$ denote the ground- and excited-state populations, respectively, while $c$ encodes the coherence between the two states. The amplitude-damping channel with Kraus operators

\begin{equation}
K_0=\begin{pmatrix}1&0\\0&\sqrt{1-\gamma}\end{pmatrix},\qquad
K_1=\begin{pmatrix}0&\sqrt{\gamma}\\0&0\end{pmatrix},
\label{eq:kraus}
\end{equation}

\noindent acts on $\rho$ as

\begin{equation}
\Lambda(\rho)
=\sum_i K_i \rho K_i^{\dagger}=
\begin{pmatrix}
p_0 + \gamma p_1 & \sqrt{1-\gamma}\,c \\
\sqrt{1-\gamma}\,c^\ast & (1-\gamma)\,p_1
\end{pmatrix}.
\label{eq:AD_action}
\end{equation}

\noindent When using dynamical decoupling, this interval is split into two equal halves of duration $\tau = dt/2$. Considering a particular dynamical decoupling sequence, the time duration can be separated by a $Y$ pulse, followed by another $Y$ (so that the net unitary is identity). For simplicity, we assume the $Y$ gates are applied instantaneously. The amplitude-damping parameter per half-interval becomes
\begin{equation}
\gamma = \frac{dt}{2T_1}.
\label{eq:gamma'}
\end{equation}

\begin{figure*}[h]
 \centering
\includegraphics{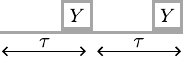}
\end{figure*}

 \noindent After a single cycle of the Amplitude Damping---$Y$---Amplitude Damping---$Y$ sequence as illustrated above, the density matrix evolves as:

\begin{equation}
\rho \mapsto \rho_{\text{final}}(\tau)
=
\begin{pmatrix}
(1-\gamma)\,p_0+ \gamma p_1 & (1-\gamma)\,c \\
(1-\gamma)\,c^\ast & \gamma p_0 + (1-\gamma)\,p_1
\end{pmatrix} + \mathcal{O}(\gamma^2).
\label{eq:single-YADY}
\end{equation}

\noindent Passing to the limit of a large number of dynamical decoupling cycles in time $t\gg dt$, the differential equation for the ground-state population is
\begin{equation}
\frac{dp_0}{dt} = \frac{1}{2T_1}\,(1 - 2p_0),
\label{eq:dp0_dt}
\end{equation}

\noindent With the initial condition $p_{0}(0)$ at $t=0$, and denoting the ground-state population at time $t$ as $p_{0}(t)$, the solution to the differential equation is

\begin{equation}
p_0(t) = e^{-t/T_1}\big(p_0(0)-\tfrac{1}{2}\big)+\tfrac{1}{2},
\qquad
p_1(t) = e^{-t/T_1}\big(p_1(0)-\tfrac{1}{2}\big)+\tfrac{1}{2}.
\label{eq:pops_solution}
\end{equation}

\noindent The coherence term decays as

\begin{equation}
c(t) = c(0)\,e^{-t/(2T_1)}.
\label{eq:coh_decay}
\end{equation}

Application of the resulting many-cycle channel to the four density matrix components of an $N$-qubit GHZ state with idle times $t_i$ on the $i^{\text{th}}$ qubit yields:
\begin{align}
\ket{0}^{\otimes N}\!\bra{0}^{\otimes N} &\mapsto 
\prod_{i=0}^{N-1} \tfrac{1}{2}\big(1+e^{-t_i/T_1}\big)\,\ket{0}^{\otimes N}\!\bra{0}^{\otimes N}
+ \prod_{i=0}^{N-1} \tfrac{1}{2}\big(1-e^{-t_i/T_1}\big)\,\ket{1}^{\otimes N}\!\bra{1}^{\otimes N}, 
\label{eq:map_00} \\[6pt]
\ket{1}^{\otimes N}\!\bra{1}^{\otimes N} &\mapsto 
\prod_{i=0}^{N-1} \tfrac{1}{2}\big(1-e^{-t_i/T_1}\big)\,\ket{0}^{\otimes N}\!\bra{0}^{\otimes N}
+ \prod_{i=0}^{N-1} \tfrac{1}{2}\big(1+e^{-t_i/T_1}\big)\,\ket{1}^{\otimes N}\!\bra{1}^{\otimes N}, 
\label{eq:map_11} \\[6pt]
\ket{0}^{\otimes N}\!\bra{1}^{\otimes N} &\mapsto 
\prod_{i=0}^{N-1} e^{-t_i/(2T_1)}\,\ket{0}^{\otimes N}\!\bra{1}^{\otimes N}, 
\label{eq:map_01} \\[6pt]
\ket{1}^{\otimes N}\!\bra{0}^{\otimes N} &\mapsto 
\prod_{i=0}^{N-1} e^{-t_i/(2T_1)}\,\ket{1}^{\otimes N}\!\bra{0}^{\otimes N}. 
\label{eq:map_10}
\end{align}

\noindent Thus, the transformed density matrix, after discarding terms orthogonal to the GHZ state, is

\begin{align}
{\rho}_{\mathrm{GHZ}} = \frac{1}{2}\Bigg[
&\frac{1}{2^N}\Bigg(\prod_{i=0}^{N-1} \big(1+e^{-t_i/T_1}\big)
+ \prod_{i=0}^{N-1} \big(1-e^{-t_i/T_1}\big)\Bigg)\,
\big(\ket{0}^{\otimes N}\!\bra{0}^{\otimes N}+\ket{1}^{\otimes N}\!\bra{1}^{\otimes N}\big) \notag \\[6pt]
&+ \prod_{i=0}^{N-1} e^{-t_i/(2T_1)}\,
\big(\ket{0}^{\otimes N}\!\bra{1}^{\otimes N}+\ket{1}^{\otimes N}\!\bra{0}^{\otimes N}\big)
\Bigg].
\label{eq:rho_ghz}
\end{align}

\noindent The resulting fidelity is
\begin{align}
F_{T_1}
= \bra{\mathrm{GHZ}}{\rho}_{\mathrm{GHZ}}\ket{\mathrm{GHZ}}
= \tfrac{1}{2}\Big[
\tfrac{1}{2^N}\prod_{i=0}^{N-1} \big(1+e^{-t_i/T_1}\big)
+ \tfrac{1}{2^N}\prod_{i=0}^{N-1} \big(1-e^{-t_i/T_1}\big)
+ \prod_{i=0}^{N-1} e^{-t_i/(2T_1)}
\Big].
\label{eq:fidelity_T1}
\end{align}

\noindent The first two terms contribute to the $Z$-type fidelity, as they arise from the diagonal elements of the density matrix and are affected only by $X$-type Pauli errors. In contrast, the final term originates from the off-diagonal coherence and is therefore sensitive to $Z$-type Pauli errors, contributing to the $X$-type fidelity.

\subsubsection{Fidelity under Single-Qubit Gates}
\label{sssec:ghz_fidelity_sq}

To estimate the effect of single-qubit gate errors on the GHZ fidelity, we model each gate by a local depolarizing channel with Pauli error rate $e_p^{\mathrm{sq}}$. Because any nontrivial Pauli error reduces the fidelity, the contribution from $n_{\mathrm{sq}}$ number of single-qubit gates is
\begin{equation}
F_{\mathrm{sq}} = \bigl(1 - e_p^{\mathrm{sq}}\bigr)^{n_{\mathrm{sq}}}.
\label{eq:fidelity_sq}
\end{equation}

\noindent For the $Z$-type fidelity, only $X$ and $Y$ Pauli errors contribute,
\begin{equation}
F_{\mathrm{sq}}^{Z} = \bigl(1 - \tfrac{2}{3} e_p^{\mathrm{sq}}\bigr)^{n_{\mathrm{sq}}},
\end{equation}
and since the total fidelity is defined as the arithmetic mean of the $X$- and $Z$-type contributions, 
\begin{equation}
F_{\mathrm{sq}}^{X} = 2 F_{\mathrm{sq}} - F_{\mathrm{sq}}^{Z}.
\label{eq:fidelity_sq_XZ}
\end{equation}

\subsubsection{Fidelity under Two-Qubit Gates}
\label{sssec:ghz_fidelity_2q}

Similarly, let $n_{2q}$ denote the number of two-qubit gates, each followed by local depolarizing channels on the involved qubits with error rate $e_p^{\mathrm{2q}}$. Out of the $15$ non-identity two-qubit Pauli operators, all but $ZZ$ reduce the GHZ fidelity. Thus, the effective error probability is $\tfrac{14}{15}e_p^{\mathrm{2q}}$, and
\begin{equation}
F_{\mathrm{2q}} = \bigl(1 - \tfrac{14}{15}e_p^{\mathrm{2q}}\bigr)^{n_{2q}}.
\label{eq:fidelity_2q}
\end{equation}

\noindent For the $Z$-type fidelity, only $ZZ$, $IZ$, and $ZI$ errors leave the stabilizers unchanged, giving
\begin{equation}
F_{\mathrm{2q}}^{Z} = \bigl(1 - \tfrac{12}{15}e_p^{\mathrm{2q}}\bigr)^{n_{2q}}.
\end{equation}
Since the total fidelity is defined as the arithmetic mean of the $X$- and $Z$-type contributions, we obtain
\begin{equation}
F_{\mathrm{2q}}^{X} = 2F_{\mathrm{2q}} - F_{\mathrm{2q}}^{Z}.
\label{eq:fidelity_2q_XZ}
\end{equation}

\subsubsection{Fidelity under readout error}
\label{sssec:ghz_fidelity_readout}

\noindent To estimate the fidelity contribution from readout errors, we model the readout channel with Kraus operators
\begin{equation}
k_0 = \begin{pmatrix} \sqrt{1-e_0} & 0 \\[4pt] 0 & 0 \end{pmatrix}, \quad
k_1 = \begin{pmatrix} 0 & 0 \\[4pt] \sqrt{e_0} & 0 \end{pmatrix}, \quad
k_2 = \begin{pmatrix} 0 & \sqrt{e_1} \\[4pt] 0 & 0 \end{pmatrix}, \quad
k_3 = \begin{pmatrix} 0 & 0 \\[4pt] 0 & \sqrt{1-e_1} \end{pmatrix},
\end{equation}

\noindent where $e_0$ and $e_1$ denote the probabilities of misidentifying $|0\rangle$ and $|1\rangle$, respectively. 
For a qubit in state $\rho$, measurement in the $Z$ basis under this channel yields
\begin{equation}
\sum_i \Tr(k_i \rho k_i^{\dagger} Z) = \sum_i \Tr(\rho k_i^{\dagger} Z k_i) \equiv \Tr(\rho Z_\Lambda),
\end{equation}
where the transformed observable is
\begin{equation}
Z_\Lambda = (e_1-e_0)\,\mathbb{I} + (1-e_0-e_1)\,Z.
\end{equation}
By inserting basis changes before the readout channel, the corresponding transformed observables for $X$ and $Y$ are
\begin{align}
X_\Lambda &= Y^{1/2} Z_\Lambda Y^{-1/2} = (e_1-e_0)\,\mathbb{I} + (1-e_0-e_1)\,X, \\[4pt]
Y_\Lambda &= X^{-1/2} Z_\Lambda X^{1/2} = (e_1-e_0)\,\mathbb{I} + (1-e_0-e_1)\,Y.
\end{align}
The expectation value of any stabilizer is obtained by replacing each Pauli operator with its corresponding transformed observable. The resulting fidelity under asymmetric readout error, incorporating the effect on the expectation values of the $X$- and $Z$-type stabilizers of the GHZ state, is

\begin{equation}
F_{\mathrm{readout,asym}} = \tfrac{1}{4}\Big[(1-e_0)^N + e_0^N + (1-e_1)^N + e_1^N\Big] 
+ \tfrac{1}{2}(1-e_0-e_1)^N + \tfrac{1}{2^{N/2}}(e_1-e_0)^N \cos\!\left(\tfrac{\pi N}{4}\right).
\end{equation}
Since in practice $e_0 \approx e_1 \lesssim 1\%$, as shown in Fig.~\ref{sfig:error_rates}, it is convenient to approximate the asymmetric model by the symmetric case $e_0=e_1=\varepsilon$. Under this assumption, the total fidelity simplifies to

\begin{equation}
F_{\mathrm{readout,sym}} = \tfrac{1}{2}\Big[(1-2\varepsilon)^N + (1-\varepsilon)^N + \varepsilon^N\Big].
\label{eq:sym_ro_fid}
\end{equation}
The corresponding fidelities for $X$-type and $Z$-type stabilizers are
\begin{equation}
F_{\mathrm{readout,sym}}^X = (1-2\varepsilon)^N, \qquad 
F_{\mathrm{readout,sym}}^Z = (1-\varepsilon)^N + \varepsilon^N.
\label{eq:sym_ro_x_fid}
\end{equation}
Finally, we approximate all error channels as independent. In our framework, the fidelity is estimated
post measurement and therefore includes readout errors, consistent with the conditions of the GHZ game.
The combined fidelity is bounded by the product of contributions from the independent error channels,
\begin{equation}
F_{\mathrm{total}} = F_{T_1} \times F_{\mathrm{sq}} \times F_{\mathrm{2q}} \times F_{\mathrm{readout,sym}}.
\label{eq:total_fid}
\end{equation}

\begin{figure*}[b!!]
\centering
\includegraphics{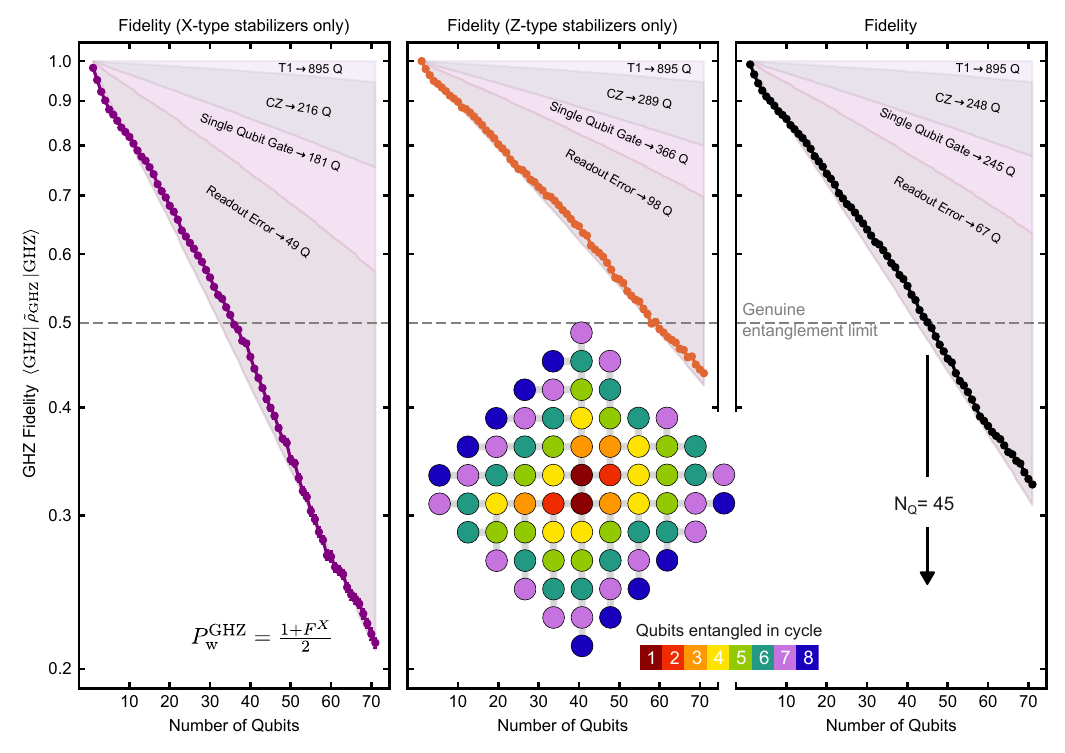}
\caption{Experimental and estimated $X$-type, $Z$-type, and total fidelities of the GHZ state as a function of system size, up to 71 qubits. Shaded regions indicate the contributions of individual error channels, with mean values used for all estimates (see \cref{sec:ghz_stab_error_budget}): 
single-qubit Pauli error $e_p^{\mathrm{sq}} = 5\times 10^{-4}$, 
two-qubit Pauli error $e_p^{\mathrm{2q}} = 3\times 10^{-3}$, 
symmetric readout error $\varepsilon = 7\times 10^{-3}$, 
and relaxation time $T_1 = 73~\mu\mathrm{s}$.
 The measured total fidelity drops below 0.5 at $N=45$, which serves as a witness of genuine $N$-partite entanglement (the genuine entanglement limit).
 Labels within the shaded regions mark the system size at which each channel alone would reduce the fidelity below 0.5. The shaded wedge corresponding to single-qubit errors is relatively large due to the additional single-qubit gates introduced by dynamical decoupling. Inset: representative two-dimensional GHZ growth pattern, with colors denoting the entangling-cycle index at which qubits are added to the GHZ state.}

\label{sfig:ghz_fidelity_error_estimate}
\end{figure*}

\newpage 

\subsection{GHZ State Growth and Fidelity}
\label{ssec:further_data_ghz_game}

The growth of an $N$-qubit GHZ state is implemented using a randomized breadth-first search (BFS)\,\cite{BFS_cormen2022} on a two-dimensional qubit grid. The procedure begins with a single qubit and expands outward by sequentially entangling neighboring qubits. At each step, the set of unentangled qubits adjacent to the GHZ state is identified, and the order of their inclusion is randomized by shuffling this set. This stochastic growth path distributes the effect of local noise and device-specific imperfections. Averaging results over multiple randomizations yields spatially homogeneous statistics. The entangling CNOT operation is realized with a gate sequence $\{\text{Hadamard},\text{CZ},\text{Hadamard}\}$ on the target qubit, which incorporates a new qubit into the existing GHZ state. Owing to the radial expansion on a two-dimensional geometry, the required two-qubit gate depth scales approximately as $\sqrt{N}$.

For each system size, we generate 50 randomized growth patterns and, for each pattern, perform 20 independent instances of the GHZ game (choices of the $x_i$). This results in $N_{g}=1000$ games per system size, with each game executed $N_{s}=1000$ times. In each shot, a win is recorded when the parity of the measured output bitstring fulfills the GHZ winning condition defined in the main text. The mean winning probability, averaged across the 50 randomizations, is shown in Fig.~3 of the main text. The standard error of the mean is given by

\begin{equation}
\sigma_{N} = \frac{SD_{50}}{\sqrt{50}},
\label{eq:ghz_sigma}
\end{equation}
where $SD_{50}$ denotes the standard deviation of the average win rate across the 50 growth patterns. The winning probability of the GHZ game, $P_{\mathrm{w}}^{\mathrm{GHZ}}$, is directly related to the $X$-type stabilizer fidelity, $F^{X}$, through 
$P_{\mathrm{w}}^{\mathrm{GHZ}} = (1+F^{X})/2$. 
The total GHZ fidelity is obtained as the average of the $X$- and $Z$-type stabilizer fidelities. To obtain both contributions, $Z$-type measurements are interleaved with the game experiments, and the order of all circuits is randomized. Error suppression is implemented through dynamical decoupling and randomized compiling (one instance per $X$-type stabilizer circuit). Fig.~\ref{sfig:ghz_fidelity_error_estimate} summarizes the measured $X$-, $Z$-, and total fidelities, together with an error analysis described in \cref{sec:ghz_stab_error_budget}.

\newpage

\newpage

\vspace{300mm}

\section{Shallow circuits and Hidden Linear Function (HLF) Problem}
\label{sec:shallow circuits}

\subsection{Formalism and Further Experimental Data}
\label{ssec:further_data_hlf}

\begin{figure*}[htb!]
\centering
\includegraphics{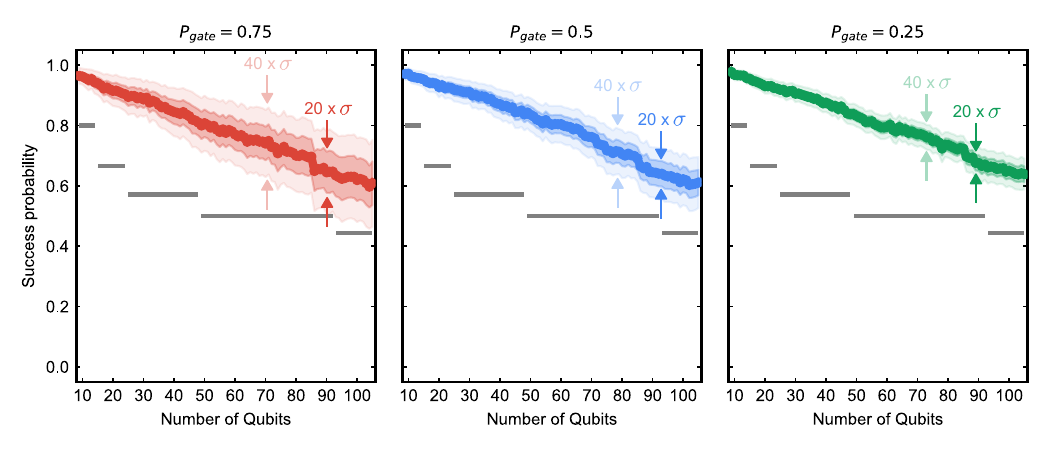}
\caption{Additional data for the shallow-circuit experiment at different gate application 
probabilities. The trends indicate that similar success probability is obtained in solving the HLF problem at a given dimension. Shaded regions denote confidence intervals: dark shading 
corresponds to $\pm 10\sigma$ and light shading to $\pm 20\sigma$. Numerical studies up to a few tens 
of qubits indicate that an extensive set of bitstrings represents valid solutions for a given problem, 
so that even in the presence of errors, the measured output bitstring may still yield a correct solution.
}
\label{sfig:shallow_circuits_full_data}
\end{figure*}

Our work experimentally implements the two-dimensional Hidden Linear Function (HLF) problem\,\cite{bravyi2018quantum}. We note that the theoretical results of Ref.\,\cite{bravyi2018quantum} establish a quantum--classical separation for circuits achieving correctness with probability $p = 7/8$, whereas our implementation operates in the regime of exact solutions.

We conjecture that the minimum classical circuit depth required to implement a Boolean function, using two-input, one-output gates, that depends on all $s$ inputs and solves the HLF problem, is $\log_{2}(s)$. To encode the two-dimensional Hidden Linear Function 
(HLF) problem\,\cite{bravyi2018quantum} on a square qubit grid, the diagonal entries of the defining 
matrix $A$ are mapped to S gates on individual qubits, while the off-diagonal entries correspond to 
CZ gates between qubits. Pairs of qubits that are not connected in the two-dimensional geometry always 
correspond to zero entries in $A$. An illustrative example of mapping an $n \times n$ matrix $A$, with $n=9$, onto a grid is shown below.

\begin{figure*}[h]
 \centering
\includegraphics{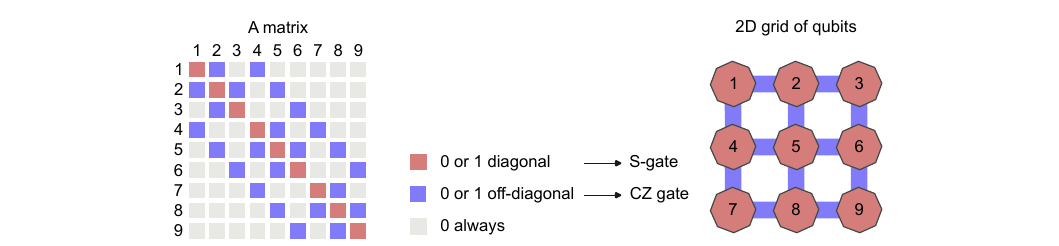}
\end{figure*}

\noindent Because all Boolean inputs of the 2D HLF problem acquire a geometric representation, the total number of inputs can be counted directly from the grid. We conjecture that a lower bound on the number of classical circuit layers to exactly solve the HLF problem with total number of inputs $s$ = (number of edges + number of nodes) is given by
\begin{equation}
 L_{\mathrm{classical}}=\log_{2}(s),
 \label{eq:shallow_circuits_classical_bound}
\end{equation}
\noindent where the terms \textit{edges} and \textit{nodes} refer to the elements of the grid's graph. In the 3x3 example schema above, $s = 9 + 12 = 21$. This expression provides a measure of classical computational complexity for comparison with the quantum implementation.

For the quantum case, the target unitary is implemented using $S$ gates together with four layers of CZ gates. 
Fig.~4 shows representative 2D HLF instances, constructed by randomly applying gates to the allowed geometry 
with probability $P_{\rm gate}=0.5$ and evaluated in terms of the success probability. 
Additional data are provided in Fig.~\ref{sfig:shallow_circuits_full_data}, including gate application probabilities 
of $P_{\rm gate}=0.25$ and 0.75 for both $S$ and CZ gates, complementing the results in Fig.~4. For each system size, 
the mean value is computed as the average over $N_{g}=1000$ random HLF instances, each executed $N_{s}=100$ times. 
The standard error of the mean is obtained from the distribution of results across the same ensemble and is defined as

\begin{equation}
\sigma_{N} = \frac{SD_{1000}}{\sqrt{1000}},
 \label{eq:shallow_sigma}
\end{equation}
with $SD_{1000}$ denoting the standard deviation across the ensemble of 1,000 HLF problems.

\end{document}